\documentclass{jfm}
\usepackage{graphicx}
\usepackage{epstopdf, epsfig}
\usepackage{xcolor}
\usepackage{bm}
\usepackage{physics}
\usepackage{tabularx}
\usepackage{wasysym}
\usepackage{tikz}

\shorttitle{Particle Segregation Force}
\shortauthor{Y. Duan, L. Jing, P. B. Umbanhowar, J. M. Ottino, and R. M. Lueptow}

\title{General model for segregation forces in flowing granular mixtures}

\author{Yifei Duan\aff{1},
 Lu Jing\aff{2},
 Paul B. Umbanhowar\aff{3},
 Julio M. Ottino\aff{1,3,4},
\and Richard M. Lueptow\aff{1,3,4}
 \corresp{\email{r-lueptow@northwestern.edu}}
}
\affiliation{\aff{1}Department of Chemical and Biological Engineering, Northwestern University, Evanston, Illinois 60208, USA
\aff{2}Institute for Ocean Engineering/Water Research Center, Shenzhen International Graduate School, Tsinghua University, Shenzhen 518055, China\
\aff{3}Department of Mechanical Engineering, Northwestern University, Evanston, Illinois 60208, USA
\aff{4}Northwestern Institute on Complex Systems (NICO), Northwestern University, Evanston, Illinois 60208, USA
}

\begin{document}

\maketitle

\begin{abstract}
Particle segregation in dense flowing size-disperse granular mixtures is driven by gravity and shear, but predicting the associated segregation force due to both effects has remained an unresolved challenge. Here, a model of the combined gravity- and kinematics-induced segregation force on a single intruder particle is integrated with a model of the concentration dependence of the gravity-induced segregation force. The result is a general model of the net particle segregation force in flowing size-bidisperse granular mixtures. Using discrete element method simulations for comparison, the model correctly predicts the segregation force for a variety of mixture concentrations and flow conditions in both idealized and natural shear flows. 
\end{abstract}

\section{Introduction}
Particle segregation in flowing granular materials has significant implications for flow mobility, rheology, and mixing, a fact reflected in the extensive attention given to this topic in the fields of granular flow mechanics, geophysical flows, and chemical engineering processes~\citep{ottino_mixing_2000,ottino_mixing_2008,frey_how_2009,johnson_grain-size_2012}. Recent advances in continuum advection-diffusion-segregation models allow successful prediction of segregation in canonical granular flow configurations~\citep{gray_particle_2018,umbanhowar_modeling_2019,thornton_brief_2021}, although this approach requires building generalized constitutive relations for segregation. Nevertheless, the basics are straightforward: segregation in dense granular flows is driven by gravity and shear \citep{fan_phase_2011,jing_unified_2021,liu_coupled_2023,singh_continuum_2023}. 

Gravity-induced segregation is generated by percolation of small particles through voids between large particles~\citep{savage_particle_1988} and buoyancy effects whereby heavier particles sink relative to lighter particles~\citep{xiao_modelling_2016}. Shear-induced segregation can be driven by enhanced percolation due to kinetic sieving (shear opens voids underneath small particles)~\citep{savage_particle_1988} and migration of particles along shear gradients~\citep{fan_phase_2011}. These mechanisms can cooperate or compete depending on the forces driving the flow, leading to complex and sometimes seemingly contradictory phenomena~\citep{guillard_scaling_2016,jing_unified_2021}. For example, while large particles typically rise to the top of gravity-driven free-surface granular flows~\citep{staron_segregation_2014}, there are also conditions where large particles instead sink to the bottom of the flow~\citep{thomas_reverse_2000,felix_evidence_2004} or migrate to high shear rate regions~\citep{fan_phase_2011}.  On the flip side, there can be a benefit to these effects: size segregation and density segregation can be used strategically to offset one another to avoid segregation altogether~\citep{alonso1991optimum,duan_modelling_2021, duan_no_segreg_2023}. Although these strategies and observations have advanced our knowledge, much remains to be understood about segregation in granular flows, particularly with regard to the forces at the particle scale that lead to segregation.

Recent detailed characterization of the driving and resisting forces of segregation on individual particles or collections of particles in granular flows has informed an emerging bottom-up framework for segregation flux modeling~\citep{tripathi_theory_2021,rousseau_bridging_2021,duan_segregation_2022,sahu_particle_2023,yennemadi_drag_2023}. The essential idea is to first ascertain the segregation driving force, $F_{seg},$ and the resistive drag force, $F_{drag},$ at the particle level~\citep{guillard_scaling_2016,van_der_vaart_segregation_2018,jing_rising_2020,jing_unified_2021,jing_drag_2022,liu_lift_2021}, and then derive the stress gradients between particle species at the continuum level based on homogenization~\citep{rousseau_bridging_2021,duan_segregation_2022}. This approach can form the basis for further derivations of the segregation flux in a mixture theory framework~\citep{gray_theory_2005,rousseau_bridging_2021}. In this process, the key step is determining the functional forms of $F_{seg}$ and $F_{drag}$, as discussed extensively in recent literature. Briefly, at the single intruder limit, $F_{seg}$ consists of a gravity-induced, buoyancy-like term and a shear-gradient-induced term~\citep{guillard_scaling_2016,van_der_vaart_segregation_2018,jing_unified_2021,liu_lift_2021}, whereas $F_{drag}$ can be captured in terms of a Stokes-like drag~\citep{tripathi_numerical_2011, jing_drag_2022}. However, a complete description of how $F_{seg}$ and $F_{drag}$ depend on particle concentration remains to be developed~\citep{bancroft_drag_2021,duan_segregation_2022}. Here we focus on $F_{seg}$ and its dependence on particle concentration in flows where both buoyancy and shear-gradient effects may be present.

The segregation force is defined as the net force on a particle in the segregation direction resulting from interactions with other flowing particles, that, when combined with other forces acting on the particle (such as its weight), drives segregation when the total force is unbalanced~\citep{guillard_scaling_2016,jing_unified_2021}. Despite its simple definition, measuring $F_{seg}$ directly is challenging in physical experiments due to the small magnitude of $F_{seg}$ relative to the force fluctuations in rapidly flowing granular materials. Alternatively, although discrete element method (DEM) or other particle dynamics simulation methods can provide detailed force information on any particle in a granular mixture, direct calculation of the net contact force on freely segregating particles does not readily serve to characterize $F_{seg}$, because segregation almost always occurs in a quasi-equilibrium state where the measured net contact force (i.e., $F_{seg}+F_{drag}$) is balanced by the particle weight~\citep{tunuguntla_discrete_2016,staron_rising_2018}. To solve this problem, a ``virtual force meter" was proposed by~\cite{guillard_scaling_2016} and rapidly adapted to a variety of flows~\citep{guillard_scaling_2016,van_der_vaart_segregation_2018,jing_rising_2020,jing_unified_2021,liu_lift_2021}. This approach uses DEM simulations to consider the single intruder limit (i.e., where the volume concentration of species $i$ in a bidisperse granular mixture approaches zero, $c_i\rightarrow0$) in a ``bed" of flowing particles, which are typically smaller than the intruder particle. The single spherical intruder particle is attached to a virtual spring that acts only in the segregation direction, typically the $z$-direction, which is perpendicular to the flow in the $x$-direction. The spring constrains the intruder to remain at an average equilibrium $z$-position, but does not restrict its other degrees of freedom. Most importantly, the mean spring extension provides the spring force from which the segregation force $F_{seg}$ for a given set of simulation conditions can be found after accounting for the particle weight (or other forces). 

Using the virtual spring approach, we recently developed a model~\citep{jing_unified_2021} for $F_{seg}$ on a single intruder particle of species $i$, denoted ${F}_{i,0}  \equiv F_{seg}\Big|_{c_i\rightarrow0}$, 
which has been validated in various free-surface and wall-confined granular flows. This single intruder segregation force model has two terms, one related to gravity and the other related to flow kinematics:
\begin{equation}
    \label{eq:fseg_org}
    F_{i,0}=-f^g(R)\frac{\partial{p}}{\partial{z}}V_i+f^k(R)\frac{p}{\dot\gamma}\frac{\partial\dot\gamma}{\partial{z}}V_i,
\end{equation}
\noindent where superscripts $g$ and $k$ indicate gravity- and kinematics-related mechanisms, respectively, the dimensionless functions $f^g(R)$ and $f^k(R)$ depend on the intruder-to-bed-particle size ratio $R$ (expressions are provided later in equation (\ref{eq:intruder})), $V_i$ is the intruder volume, $p$ is the pressure, $\dot\gamma$ is the local shear rate, and $\rho$ is the density of both the intruder and the bed particles. 
Here, ``pressure" and vertical ``normal stress" are interchangeable ($p\equiv\sigma_{zz}$) under the assumptions that granular flows at steady state are incompressible and the deviatoric stress aligns with the strain rate tensor~\citep{kim2023second}. In dimensionless form, (\ref{eq:fseg_org}) can be expressed as:

\begin{subequations}
\begin{equation}
  \label{eq:fseg_breve}
  \hat{F}_{i,0}=\hat{F}^g_{i,0} + \hat{F}^k_{i,0},
\end{equation}
where the hat diacritic ($~\hat{}~$) denotes dimensionless forces scaled for reference by particle weight, $m_ig_0$, in Earth gravity, $g_0=9.81$\,m/s$^2$.  Accordingly, the normalized gravity- and kinematics-induced segregation forces on a lone intruder particle are
\begin{equation}
        \hat{F}^g_{i,0}=-f^g(R)\frac{\partial p}{\partial z} \frac{1}{\rho g_0},
    \label{eq:fseg1}
\end{equation}
\begin{equation}
        \hat{F}^k_{i,0}=f^k(R)\frac{\partial \dot\gamma}{\partial z} \frac{p} {\dot\gamma \rho g_0}.
    \label{eq:fseg2}
\end{equation}
\label{eq:fseg}
\end{subequations}

\noindent Note that although here we use $g_0$ for  normalization, our models work for arbitrary values of gravitational acceleration (including $g=0$) as shown previously \citep{jing_rising_2020,jing_unified_2021}. The first term on the r.h.s.\ of (\ref{eq:fseg_breve}), $\hat{F}^g_{i,0},$ can be viewed as a gravity-dependent effective-buoyancy force, while the second term, $\hat{F}^k_{i,0},$ is a flow kinematics-based shear-rate-gradient term. We have previously shown~\citep{jing_unified_2021} that if a local rheology is assumed, (\ref{eq:fseg_breve}) is equivalent to an earlier model of similar form in which the shear stress gradient is used instead of the shear rate gradient~\citep{guillard_scaling_2016}. Although other intruder segregation force models exist, including ones related to Saffman lift~\citep{van_der_vaart_segregation_2018}, kinetic-theory~\citep{liu_lift_2021}, and granular temperature gradients~\citep{fan_theory_2011,hill_segregation_2014}, Eq.~(\ref{eq:fseg}) is the only model to be thoroughly validated over a range of three-dimensional flow configurations including confined wall-driven flows and free surface gravity-driven flows.

\begin{figure} \centerline{\includegraphics[width=\columnwidth]{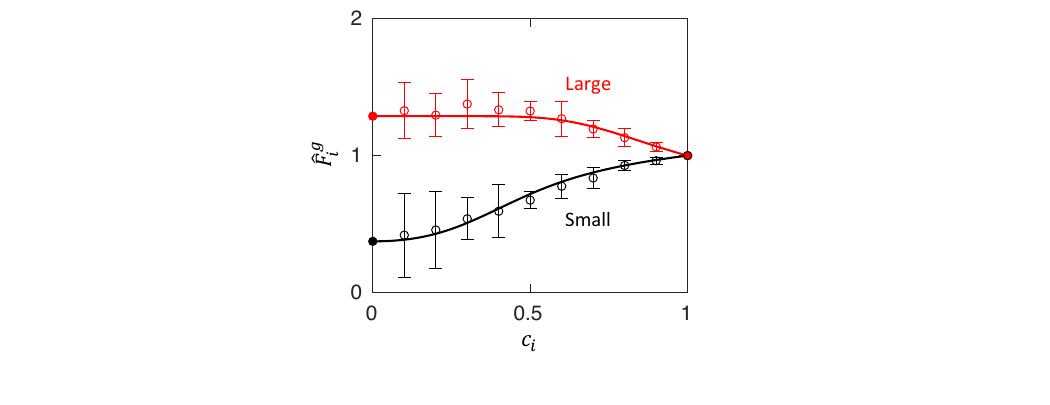}}
\caption{
Example of gravity-induced segregation force scaled by particle weight $\hat F_i^g=F_i^g/m_ig_0$ vs.\ species concentration $c_i$ for large and small particles with size ratio $R=d_l/d_s=2$ in uniform shear flow ($\hat F^k_i=0$) with $\dot\gamma=100\,\mathrm{s}^{-1}$.  Error bars show the standard deviation of depth-averaged $\hat F_i^g$ from DEM simulations~\citep{duan_segregation_2022}.  Solid circles at $c_i=0$ and curves are predictions of intruder force model (\ref{eq:fseg1}) and mixture force model (\ref{eq:fseg_g}), respectively.
}
    \label{fig_tanh}
\end{figure}

The single intruder limit for the segregation force, ${F}_{i,0} = F_{seg}\Big|_{c_i\rightarrow0}$, is now relatively well studied, but ${F}_{i}=F_{seg}\Big|_{c_i\in (0,1)}$ on a single particle in a mixture of particles with an arbitrary value of $c_i$ between $0$ and $1$ is much less understood, although linear and quadratic dependencies of ${F}_{i}$  on $c_i$ have been previously assumed~\citep{rousseau_bridging_2021,tripathi_theory_2021}. To explore the segregation force at finite concentrations, we recently extended the virtual spring approach for a single intruder particle~\citep{guillard_scaling_2016} to size-bidisperse mixtures of arbitrary species concentration~\citep{duan_segregation_2022} in order to characterize the dependence of the gravity-induced portion of the segregation force, $F_i^g$, on $c_i$ in a controlled horizontal uniform-shear flow (i.e., $\partial\dot\gamma/\partial{z}=0$). An example of the concentration dependence of $F_i^g$ at large-to-small particle size ratio $R=d_l/d_s=2$ is shown in figure~\ref{fig_tanh} for a DEM simulation of plane shear flow (as described in section~\ref{Method}). Consistent with previous results~\citep{duan_segregation_2022}, data points for $\hat{F}_i^g=F_{i,0}^g/m_i g_0$ approach the single intruder limit, $\hat{F}_{i,0}^g$, as $c_i\rightarrow0$. For large particles, $\hat{F}_{l,0}^g>1$ as $c_i\rightarrow0$, indicating that the upward segregation force exceeds the particle weight, resulting in a tendency for a large particle to rise; for small particles, $\hat{F}_{s,0}^g<1$ as $c_i\rightarrow0$, indicating that the upward segregation force is less than the particle weight, resulting in a tendency for a small particle to sink. At $c_i=1$, $\hat{F}_{i,0}^g=1$ for both large and small particles, indicating that the segregation force equals the particle weight such that no segregation occurs as required for the monodisperse case. 

A semi-empirical model~\citep{duan_segregation_2022} can be used to express the concentration-dependent gravity-induced segregation force on particles of species $i,$ $F_i^g,$ in terms of the gravity-induced segregation force on a single intruder particle, ${F}^g_{i,0}$, and the small and large particles concentrations, $c_s$ and $c_l$, respectively, such that for a large particle
\begin{subequations}
\begin{equation}
    \label{eq:fseg_g_large}
    \hat F_l^g=1+(\hat{F}^g_{l,0}-1)\textrm{tanh}\Big( \frac{1-\hat{F}^g_{s,0}}{\hat{F}^g_{l,0}-1}\frac{c_s}{c_l} \Big),
\end{equation}
where $\hat{F}^g_{l,0}$ and $\hat{F}^g_{s,0}$ are the gravity-induced dimensionless segregation force on a small or large intruder particle, respectively, and $c_s+c_l=1$. The analogous equation for a small particle is 
\begin{equation}
    \label{eq:fseg_g_small}
   \hat{F}^g_{s}=1-(\hat{F}^g_{l,0}-1){\frac{c_l}{c_s}}\textrm{tanh}\Big( \frac{1-\hat{F}^g_{s,0}}{\hat{F}^g_{l,0}-1}\frac{c_s}{c_l} \Big).
\end{equation}
\label{eq:fseg_g}
\end{subequations}

Equation (\ref{eq:fseg_g}) fits the data in figure~\ref{fig_tanh} quite well. The empirical hyperbolic tangent form of (\ref{eq:fseg_g}) is useful because it saturates at extremes of the domain, as is the case here where $\hat{F}_i^g$ approaches the single intruder limit as $c_i\rightarrow0$ and approaches the monodisperse limit of 1 as $c_i\rightarrow1$. Note that the asymmetry between segregation forces for large and small particles leads to different expressions for the two species.
Furthermore, the two equations in (\ref{eq:fseg_g}) depend only on the segregation force on small and large intruder particles, $\hat{F}^g_{s,0}$ and $\hat{F}^g_{l,0}$, and the concentration of small and large particles, $c_s$ and $c_l$. No knowledge of the segregation force for $0<c_i\leq1$ is needed. Moreover, the hyperbolic tangent dependence of the large particle segregation force (\ref{eq:fseg_g_large}) satisfies the theoretical constraints, namely that lim$_{c_l\rightarrow0}\tanh{(c_s/c_l)}=1$ and lim$_{c_l\rightarrow1}\tanh{(c_s/c_l)}=0$ such that $\hat F_l^g=\hat{F}_{l,0}^g$ at $c_l=0$ and $\hat F_l^g=1$ at $c_l=1$. Likewise for a small particle, (\ref{eq:fseg_g_small}) satisfies $\hat F_s^g=\hat{F}_{s,0}^g$ at $c_s=0$ (since $\tanh(A) \approx A$ for $A\rightarrow 0$) and $\hat F_s^g=1$ at $c_s=1$. These equations also meet the requirement that the total segregation force across both species for the entire system sums to the total particle weight, which can be expressed as~\citep{duan_segregation_2022}
\begin{equation}
    \label{eq:fseg_sum}
    c_l \hat F_l^g + c_s \hat F_s^g=1.
\end{equation}

With the concentration-dependent expression for the  gravity-driven segregation force (\ref{eq:fseg_g}) specified, the challenge at this point, and the focus of this paper, is extending the finite concentration framework to include the single intruder limit kinematics-related term in (\ref{eq:fseg2}). 
To this end, we build upon the models portrayed in (\ref{eq:fseg}) and (\ref{eq:fseg_g}) to extend this approach to  the total segregation force, $F_i$, on a particle due to both gravity-induced and kinematics-induced effects for arbitrary concentration size-bidisperse mixtures. We then validate the predictions of the full model with comparisons to DEM results from a variety of canonical granular flows. The ultimate goal is a segregation force model encompassing the full range of flow and particle conditions that can be broadly applied to a wide variety of situations.

\section{Method}
\label{Method}
 
\begin{figure}
    \centerline{\includegraphics[width=\columnwidth]{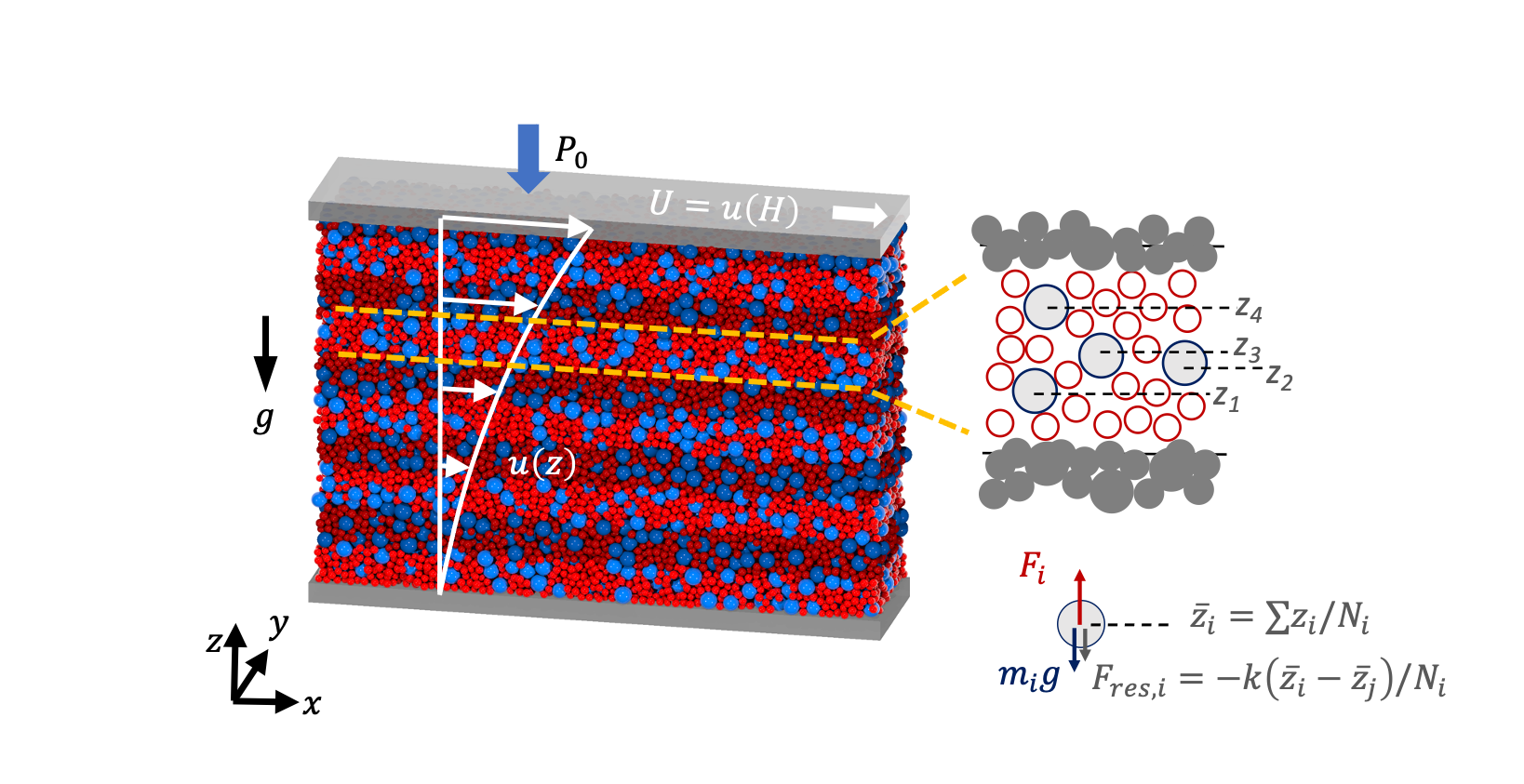}}
\caption{Large (4\,mm, blue) and small (2\,mm, red) particles ($c_l=c_s=0.5$) in a generic shear flow partitioned into $2.5d_l$ high layers (shading). Within each layer, a vertical spring-like restoring force measurement approach quantifies the average local segregation force across all particles of a particular species (small or large) in that layer. 
}
    \label{fig_fres}
\end{figure}

An in-house discrete element method DEM code running on CUDA-enabled NVIDIA GPUs is used to simulate a size-bidisperse particle mixture with species specific volume concentration $c_i$, diameter $d_i$, and density $\rho_i=1$\,g\,cm$^{-3}$ ($i=l,s$ for large or small particles, respectively). The mixture is sheared in the streamwise ($x$) direction. Boundary conditions are periodic in $x$ and $y$ with length $L=35d_l$ and width $W=10d_l$, respectively. The height is $H=25d_l$ to $50d_l$ (varied as needed) in the $z$-direction, which is normal to the flow direction. Gravity may be aligned with the $z$-direction, as shown in figure~\ref{fig_fres}, at an angle $\theta$ with respect to $z$ for inclined chute flow, or parallel to the flow aligned with $x$ for vertical chute flow. In some cases, gravity is set to zero. The standard linear spring-dashpot model~\citep{cundall1979discrete} is used to resolve particle-particle and particle-wall contacts of spherical particles using a friction coefficient of $\mu=0.5$, a restitution coefficient of 0.9, and a binary collision time of 0.15\,ms. We have confirmed that our results are relatively insensitive to these values except for very low friction coefficients ($\mu\le0.2$)~\citep{duan2020segregation,jing_rising_2020}. Large ($d_l=4$\,mm) and small ($d_s$ varied to adjust the size ratio, $R=d_l/d_s$) particle species have a $\pm10$\% uniform size distribution to minimize layering~\citep{staron_segregation_2014} (increasing the size variation to $\pm20$\% does not alter the results).  
From 26000 to 150000 particles are included in each simulation depending on the value of $R$.

The modified virtual spring approach used to measure $F_i$ in finite concentration uniform shear flows~\citep{duan_segregation_2022} must be further modified for flows with shear rate gradients since, as (\ref{eq:fseg2}) indicates, the kinematic term can be depth dependent through the pressure (depending on $g$), the shear rate gradient, or both. For the uniform shear flow method used previously to measure $F_i$~\citep{duan_segregation_2022}, a spring-like vertical restoring force proportional to the displacement in the vertical centre of mass positions for each of the two initially mixed species is applied uniformly to all particles of each species at each simulation time step. From this restoring force, the average value of $F_i$ for each species can be determined based on the average vertical displacement of each species and the applied spring constant. Not only does this allow the measurement of $F_i$, but it also simultaneously suppresses segregation throughout the flow domain, which otherwise would change the local species concentration.  In the variation of this approach used here for depth varying segregation forces, the flow domain is partitioned into layers normal to the segregation direction that are $2.5d_l$ ($1\,$cm) thick (alternating shaded and unshaded regions in the $H=25d_l$ deep bed in figure~\ref{fig_fres}). Particles are assigned to their corresponding layer and remain part of that layer regardless of their subsequent vertical displacement. At each time step a layer-specific spring-like vertical restoring force is uniformly applied to each particle of species $i$ in the layer, $F_{\mathrm{res},i}=-k( \bar z_{i}-\bar z_j)/N_i$, where the centre of mass of species $i$ is $\bar z_{i}={\sum_{p\in i}^{N_i} z_p}/N$, subscript $j$ indicates the other species, and $N_i$ and $N$ are the number of particles of species $i$ and the total number of particles in the layer, respectively. 
Note that in each layer the applied restoring forces balance, i.e., $F_{\mathrm{res},i} N_i+F_{\mathrm{res},j}N_j=0$, and the bulk flow behavior (e.g., shear flow, bulk pressure) is  unaltered. 
The spring constant is typically $k=100\,$N/m, although results are not sensitive to $k$~\citep{jing_unified_2021, duan_segregation_2022}. As shown in the free body diagram for a large particle in the lower right of figure~\ref{fig_fres}, the segregation force, $F_i$, is determined from the magnitude of the restoring force after accounting for the weight of the particle due to gravity, $g_0$, or the component of gravity in the $z$ direction, $g_z=g_0\cos{\theta}$. %This approach allows us to characterize the depth-varying particle segregation force for each species in flowing bidisperse mixtures of arbitrary concentration. 

The advantages of the restoring force measurement approach lie in its ability to suppress overall particle segregation and characterize depth-varying segregation forces while simultaneously allowing individual particles to move freely. However, collisional diffusion results in some particles dispersing outside their initially assigned measurement layers, which may corrupt the segregation force measurement when the segregation force varies with depth. Additionally, species-dependent differences in diffusion rates can potentially affect the force balance measurement approach for particles close to the boundaries. Consider, for example, a uniform flow with $\partial p/\partial z=0$ and $\partial \dot\gamma/\partial z=0$ such that there are no segregation forces.  If large particles assigned to a wall-adjacent layer diffuse away from the wall more rapidly than small particles in the same layer, the resulting increase in the centre of mass position difference between the two species will produce an associated restoring force. To quantify the potential effects of diffusion on the measured segregation forces, results for particles assigned to layers based on their initial vertical positions are compared to results where the layer assignment occurs at the start of the measurement averaging interval, which varies from 3 to 30\,s after shear onset for the various flow conditions. The latter approach ensures that diffusion over a relatively short averaging interval (2\,s) is insignificant. The overall differences in these two approaches is less than 5\%  in all cases, indicating that particle diffusion has minimal impact on the restoring force measurement approach even near the boundaries. Nevertheless, to minimize the potential effects of diffusion on force measurement, we initially assign particles to the vertical layers at shear onset to prevent segregation and then re-assign the particles to their current layers at the start of the measurement averaging interval.

\begin{figure}
    \centerline{\includegraphics[width=\columnwidth]{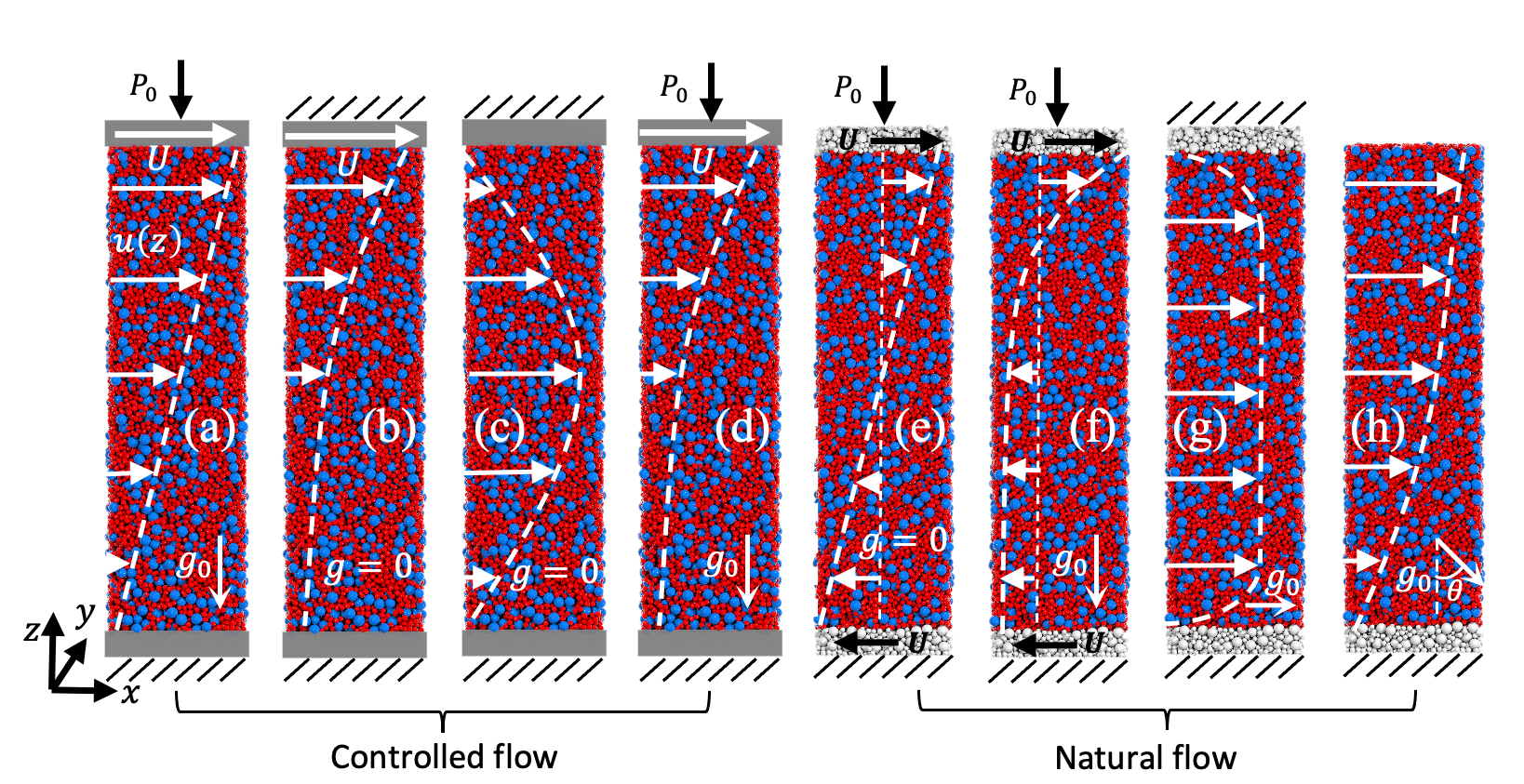}}
\caption{Schematics of flow configurations studied here (streamwise length shown is 1/5$^\mathrm{th}$ the simulated length) with periodic streamwise ($x$) and spanwise ($y$) boundaries and vertical wall boundaries as indicated (no friction/rough, fixed vertical position/pressure $P_0$, streamwise stationary/moving).  Controlled shear flows with prescribed (a) linear velocity profile with gravity; (b) exponential velocity profile without gravity; (c) parabolic velocity profile without gravity; (d) exponential velocity profile with gravity. Natural flows: (e) wall-driven without gravity; (f) wall-driven with gravity; (g) vertical chute with gravity; and (h) inclined chute with gravity. Walls with hash marks do not move vertically.
}
    \label{fig2}
\end{figure}

The force measurement approach illustrated in figure~\ref{fig_fres} is applied to a variety of flow configurations, including controlled shear flows and natural uncontrolled flows, each of which is shown schematically in figure~\ref{fig2}. For the controlled shear flows shown in figure~\ref{fig2}(a-d), a stabilizing algorithm (explained below) enforces a prescribed velocity profile between the two geometrically smooth, frictionless horizontal walls. 
By imposing a specific velocity profile, we control the shear rate and shear rate gradient, which, according to (\ref{eq:fseg2}), play direct roles in determining $\hat{F}_{i,0}$. The presence of gravity, figure~\ref{fig2}(a, d), results in a pressure gradient in $z$, which also influences $\hat{F}_{i,0}$ by virtue of both (\ref{eq:fseg1}) and (\ref{eq:fseg2}). 

Three controlled-velocity profiles are investigated: $u=Uz/H$ (linear), $Ue^{k(z/H-1)}$ (exponential), and $4U(z/H-z^2/H^2)$ (parabolic). The linear velocity profile corresponds to ideal uniform shear flow driven by a moving wall (figure~\ref{fig2}(a)). A confining overburden pressure, $P_0$, is applied to the upper wall, which is free to move vertically, and $g$ is in the $z$-direction. This flow configuration matches the flow field that provided the basis for the dependence of the gravity-induced segregation force on the mixture concentration, (\ref{eq:fseg_g}), and there is no kinematics-induced segregation, since $\partial \dot\gamma/\partial z = 0$. The exponential velocity profile is an idealization of free surface flow down a heap~\citep{fan_kinematics_2013}, except with an upper bounding wall and without gravity (figure~\ref{fig2}(b)) in order to focus on kinematics-induced segregation. Likewise, the parabolic velocity profile (figure~\ref{fig2}(c)), which is an idealization of vertical chute flow, has only kinematics-induced segregation. Since gravity does not contribute to the segregation force in a vertical chute, we set $g=0$ so the segregation force is a consequence of only the imposed parabolic velocity. We also consider a second version of an exponential velocity profile, except with a confining pressure, $P_0$, and a gravitational field in $z$ (figure~\ref{fig2}(d)) to examine combined gravity-induced and kinematics-induced segregation. 

In controlled shear flows (figures~\ref{fig2}(a-d)), a specified velocity profile, $u(z)$, is achieved by applying a small streamwise stabilizing force $k_v [\,u(z)-u_p(z_p)]\,$ to each particle at each DEM simulation time step to maintain the desired velocity profile, where $u_p$ and $z_p$ are the instantaneous particle velocity and position, respectively, and $k_v$ is a gain parameter~\citep{lerner_unified_2012, clark_critical_2018, fry_effect_2018, jing_rising_2020,jing_unified_2021, jing_drag_2022}.  For the two controlled-pressure cases with gravity-induced pressure gradients, figure~\ref{fig2}(a, d), and based on a recent analysis~\citep{jing_drag_2022}, we vary $k_v$ from 0.01 kg/s at the top of the bed to 0.03 kg/s at the bottom to account for the gravitational pressure gradient while avoiding altering the granular flow rheology and ensuring the desired velocity profile. %In these two cases, an overburden pressure, $P_0$, is applied to the top wall, which is free to move in the $z$-direction. 
For the two controlled-volume cases with $g=0$ and uniform pressure (figure~\ref{fig2}(b, c)), the velocity profile is enforced with a constant $k_v=0.02$\,kg/s. In these two cases, no overburden pressure is applied and the distance between the two walls, $H$, is fixed.  Varying $k_v$ between 0.0001 and 0.1 indicates that $k_v \geq 0.01$ kg/s is necessary to maintain the imposed velocity profile. Although the walls do not drive the flow, the upper wall moves with velocity $u(H)=U$ for cases in figure~\ref{fig2}(a, b, d) and the lower wall is fixed, $u(0)=0$.  Note that in the cases with exponential velocity profiles, figure~\ref{fig2}(b, d), the imposed velocity $u$ does not go to zero at the lower wall, i.e., $u(z=0)\approx 0.1 U \neq 0$. Because the imposed velocity is relatively small near the lower wall and the wall is frictionless and smooth, the finite wall-slip does not affect the results. 

To confirm that the imposed velocity fields do not unnaturally alter the results, we also consider four cases where the velocity field is not directly controlled, as shown in figure~\ref{fig2}(e-h).  The flow kinematics of these uncontrolled ``natural flows" are driven entirely by the combined effects of gravity and boundary conditions. The walls are rough in all cases, formed from a $2.5d_l$ thick layer of bonded large and small particles that move collectively. For the wall-driven flows,  figure~\ref{fig2}(e, f), an overburden pressure $P_0$ is imposed on the upper wall, which is otherwise free to move vertically, and which fluctuates by no more than ±0.05\% after an initial rapid dilatation of the particles at flow onset. Gravity results in a pressure gradient in $z$ for case (f). In both cases, the upper wall moves at velocity $u(H) = U$ in the $x$ direction and the lower wall at $u(0) = -U$ in the $-x$ direction.  With gravity, (f), the flow velocity changes rapidly with depth near the upper wall and slowly with depth near the bottom wall, while without gravity, case (e), the velocity profile varies linearly with depth as expected.  Both cases show little to no slip at either wall.  The vertical chute flow, shown in figure~\ref{fig2}(g), is driven by gravity aligned parallel to the rough bounding walls, resulting in a generally uniform velocity at the centre of the channel that goes to zero at the walls. In this case, there is no pressure gradient in $z$ to drive segregation, so any segregation in $z$ is driven by shear alone. Finally, the inclined chute flow has no upper wall (free boundary) so that particles flow due to a streamwise component of gravity, as shown in figure~\ref{fig2}(h). Here the pressure gradient in the segregation direction is $g_0 \cos{\theta}$, where $\theta$ is the inclination angle of the base (lower wall) relative to $\vec{g}$. 

For the controlled flows in figure~\ref{fig2}(a-d) and after shear onset, the particle bed dilates for 0.2-1.5\,s depending on the number of particles in the simulation and the flow configuration. After dilation ends, steady state flows with the prescribed velocity profile are achieved. A 2\,s long averaging interval for determining $F_i$ starts 3\,s after shear onset.  Reducing the averaging interval by half for the same start time increases the measurement uncertainty, but the time-averaged value of $F_i$ differs by less than 5\%. For the uncontrolled natural flows in figure~\ref{fig2}(e-h), the averaging interval for $F_i$ starts after the flows reach steady state (or quasi-steady state where flows are effectively invariant over the averaging interval), based on the spatially averaged streamwise velocity. The time to reach steady-state varies significantly depending on the geometry. The wall driven flows converge the fastest ($<5$\,s), the vertical chute flow is slower, needing about $20$\,s to reach steady state, and the inclined chute is the slowest, taking more than $30$\,s to reach steady state. For the wall driven flows, the $2$\,s averaging interval begins 7\,s after shear onset, while for the vertical and inclined chute flows it begins at 30\,s. Although reaching a steady state is necessary for minimizing uncertainties in time-averaged quantities, the model remains valid and accurately predicts the segregation force even during transient states.

\section{Segregation force model}
To predict the segregation force $F_i$ at arbitrary non-zero concentrations, it is useful to know the segregation force at zero concentration, $F_{i,0}$. The challenge in predicting $F_{i,0}$ resides in the dependence of $f^g$ and $f^k$ on the intruder-to-bed particle size ratio $R$ in the intruder force model (\ref{eq:fseg}). \cite{jing_unified_2021} provide empirical fits of $f_g$ and $f_k$ that are derived from numerous controlled-shear-flow DEM simulations:
\begin{subequations}
\begin{equation}
        f^g(R)=\bigg[ 1-c^g_1 \exp(-\frac{R}{R^g_1}) \bigg] \bigg[ 1+c^g_2\exp(-\frac{R}{R^g_2}) \bigg],
\end{equation}
\begin{equation}
        f^k(R)=f^k_\infty \bigg[ \tanh{(\frac{R-1}{R^k_1})} \bigg] \bigg[ 1+c^k_2\exp(-\frac{R}{R^k_2}) \bigg],
\end{equation}
\label{eq:intruder}
\end{subequations}\\
where $R^g_1=0.92$, $R^g_2=2.94$, $c^g_1=1.43$, $c^g_2=3.55$, $f^k_\infty=0.19$, $R^k_1=0.59$, $R^k_2=5.48$, and $c^k_2=3.63$ are fitting parameters for a variety of flow conditions. In applying these functions over a range of concentrations, we need to consider both the large particle and the small particle as the intruder in the corresponding intruder-to-bed particle size ratios of $d_l/d_s$ and $d_s/d_l$. Here we restrict our attention to size ratios of 1.5, 2,  and 3.\footnote{$f^g=2.254$ and $f^k=0.493$ for $R=1.5$, and $f^g=1.176$ and $f^k=-0.410$ for $R=1/1.5$; $f^g=2.343$ and $f^k=0.625$ for $R=2$, and $f^g=0.677$ and $f^k=-0.565$ for $R=1/2$; $f^g=2.154$ and $f^k=0.588$ for $R=3$, and $f^g=0.019$ and $f^k=-0.680$ for $R=1/3$.} 

Since (\ref{eq:fseg_g})  was developed for situations where gravity is normal to the flow direction ($g_z=g_0$), the inclined chute configuration studied here makes it necessary to account for gravity acting at an angle $\theta$ with respect to $z$. Replacing $g_0$ with $g_z=g_0 \cos \theta$, equations (\ref{eq:fseg_g}) are rewritten as
\begin{subequations}
\begin{equation}
% \begin{array}{lcl}
 % (\breve{F}^{g*}_l/m_lg_z) \tanh\big(-\frac{\breve{F}^{g*}_s /m_s g_z }{\breve{F}^{g*}_l /m_l g_z} \frac{c_s}{c_l}\big),\\
 \hat F_l^g=\cos{\theta}+(\hat{F}^g_{l,0}-\cos{\theta})\textrm{tanh}\Big( \frac{\cos{\theta}-\hat{F}^g_{s,0}}{\hat{F}^g_{l,0}-\cos{\theta}}\frac{c_s}{c_l} \Big),
% \end{array}
\label{tanh_large}
\end{equation}
\begin{equation}
 % -(\breve{F}^{g*}_l/m_l g_z) \frac{c_l}{c_s}\tanh\big(-\frac{\breve{F}^{g*}_s /m_s g_z}{\breve{F}^{g*}_l /m_l g_z} \frac{c_s}{c_l}\big).
    \hat{F}^g_{s}=\cos{\theta}-(\hat{F}^g_{l,0}-\cos{\theta}){\frac{c_l}{c_s}}\textrm{tanh}\Big( \frac{\cos{\theta}-\hat{F}^g_{s,0}}{\hat{F}^g_{l,0}-\cos{\theta}}\frac{c_s}{c_l} \Big).
\label{tanh_small}
\end{equation}
\label{tanh}
\end{subequations}\\

Here we propose and then confirm that the total segregation force at arbitrary mixture concentration and including both the gravity-induced term and the kinematics-induced term can be represented in terms of the same hyperbolic tangent relationship. Replacing $\hat{F}^{g}_i$ with $\hat{F}_i = \hat{F}^{g}_i + \hat{F}^k_i$ in (\ref{tanh_large}) and (\ref{tanh_small}) yields
\begin{subequations}
\begin{equation}
 % \breve{F}^{*}_l \tanh\big(-\frac{\breve{F}^{*}_s /m_s  }{\breve{F}^{*}_l /m_l } \frac{c_s}{c_l}\big),
 \hat F_l=\cos{\theta}+(\hat{F}_{l,0}-\cos{\theta})\textrm{tanh}\Big( \frac{\cos{\theta}-\hat{F}_{s,0}}{\hat{F}_{l,0}-\cos{\theta}}\frac{c_s}{c_l} \Big),
\label{tanh_large2}
\end{equation}
and
\begin{equation}
% \begin{array}{lcl}
    \hat{F}_{s}=\cos{\theta}-(\hat{F}_{l,0}-\cos{\theta}){\frac{c_l}{c_s}}\textrm{tanh}\Big( \frac{\cos{\theta}-\hat{F}_{s,0}}{\hat{F}_{l,0}-\cos{\theta}}\frac{c_s}{c_l} \Big).
% \end{array}
\label{tanh_small2}
\end{equation}
\label{tanh2}
\end{subequations}\\
Analogous to (\ref{eq:fseg_sum}), the total concentration-weighted segregation force across both species sums to the total particle weight in the segregation direction, which can be expressed as
\begin{equation}
    \label{eq:fseg_sum1}
    c_l \hat{F}_l + c_s \hat{F}_s=\cos\theta.
\end{equation}
Thus, the complete model for the concentration dependent particle segregation force in flows of size-bidisperse mixtures with pressure and shear rate gradients is specified by (\ref{eq:fseg}), (\ref{eq:intruder}), and (\ref{tanh2}).

\begin{figure}
    \centerline{\includegraphics[width=\columnwidth]{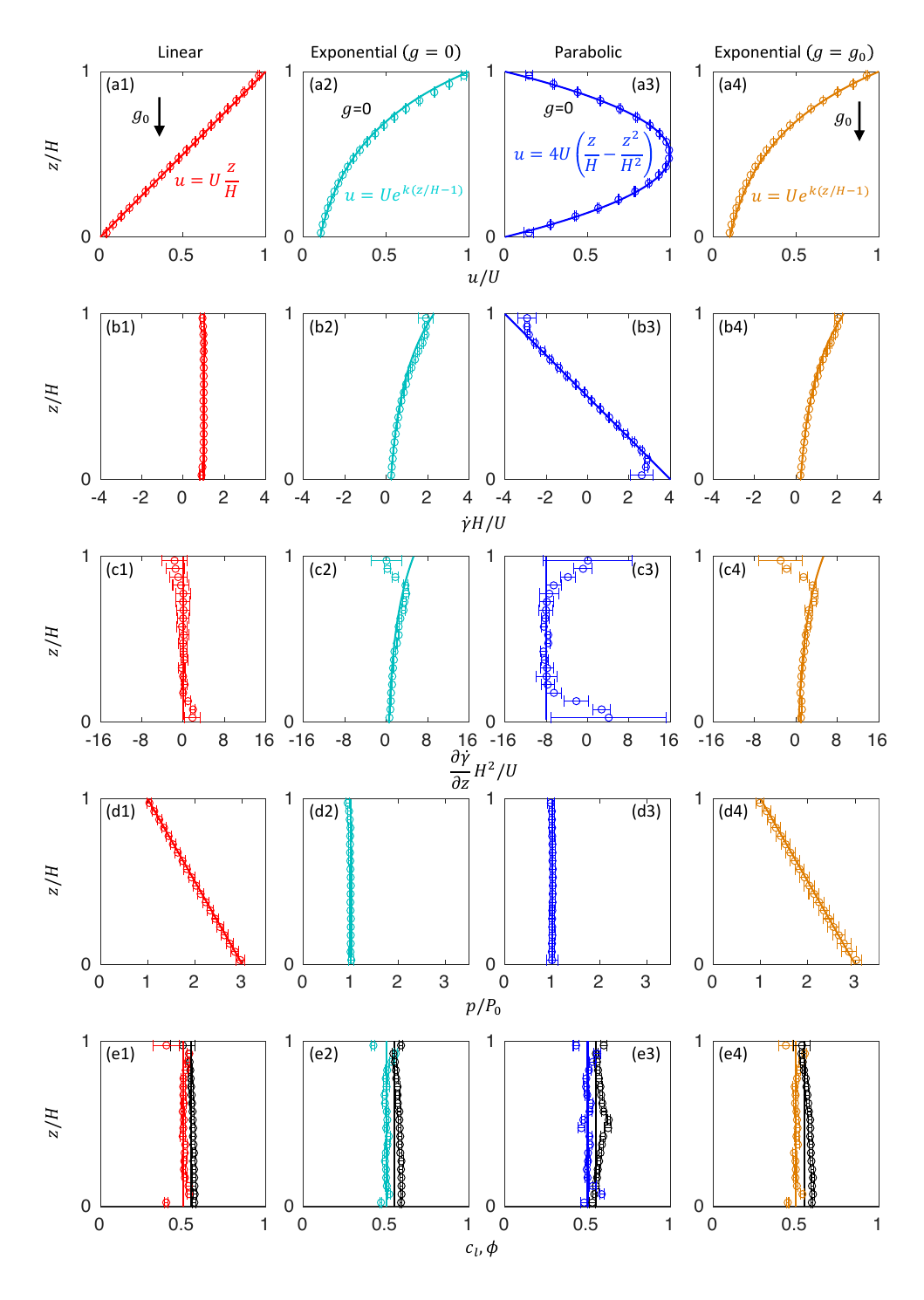}}
\caption{Scaled flow field profiles for controlled shear flows with different velocity profiles and $R=2$.  $d_l=4$\,mm, $d_s=2$\,mm, $\rho_l=\rho_s=1$\,g/cm$^3$, $H\approx 0.2\,\mathrm{m}=50d_l$, and $U=20$\,m/s. $g=g_0=9.81$\,m/s$^2$ in the negative $z$ direction for columns 1 and 4, and $g=0$ in columns 2 and 3. Values for $P_0$ are $P_0=0.5\rho \phi g_0H$ for the applied overburden pressure (columns 1 and 4), $P_0=0.61\rho \phi g_0H$ for the exponential profile (column 2), and $P_0=0.73\rho \phi g_0H$ for the parabolic profile (column 3). In the bottom row (e), $c_l$ is in color, and $\phi$ is black. 
}
    \label{fig3}
\end{figure}

\section{Results}
\subsection{Controlled shear flows}
To test the concentration-dependent particle-level segregation force model described above, i.e., (\ref{eq:fseg}, \ref{eq:intruder}, \ref{tanh2}), we first examine the controlled shear flows illustrated in figure~\ref{fig2}(a-d), as these artificial velocity profiles allow us to consider the gravity-induced and kinematics-induced contributions both separately and in combination. We can then evaluate the accuracy of the \emph{mixture} segregation force model predictions derived from equations~(\ref{tanh_large2}) and (\ref{tanh_small2}) by comparing them with DEM measurements in various flow configurations. 
For the controlled shear flows we use two size ratios, $R=2$ and $R=3$, and a 50:50 mixture of large and small particles ($c_l=c_s=0.5$), although other species volume concentrations are considered in Sec.~\ref{sec:concentration}.

Flow field details of the four controlled shear flows are shown in figure~\ref{fig3} for $R=2$. Results are similar for $R=3$. The imposed and measured streamwise velocity profiles are shown in row (a). The effectiveness of the control scheme for the velocity is evident in the close match between the DEM data points and the curves representing the target velocity profile. Rows (b, c) in figure~\ref{fig3} show the dimensionless shear rate and shear rate gradient, both of which contribute to the kinematics-induced portion of the segregation force in (\ref{eq:fseg1}). The DEM results (data points), based on finite differences for the $z$-gradients (central difference for the interior data points and single-sided difference for the edge data points), match the curves from the derivatives of the imposed velocity profiles except near the walls ($z/H=0$ and $z/H=1$). In the near-wall region, the DEM results deviate slightly from the imposed velocity profile, barely evident in row (a), but amplified by the higher derivatives associated with the shear rate and shear rate gradient in rows (b) and (c). The velocity profiles in (a) are chosen so that the shear rate gradients are zero in one case (linear, column 1) and non-zero in the other cases.  The exponential velocity profile (columns 2 and 4) has a non-zero shear rate and shear rate gradient, and both are nonlinear. For the imposed parabolic velocity profile (column 3), the shear rate and shear rate gradient measured from DEM simulation match the targeted linearly varying and constant value, respectively, only in the middle two-thirds of the channel, and their magnitudes are much larger than the other cases.

The flows in figure~\ref{fig3} also differ in their pressure fields, shown in row (d). It is important to note that the pressure gradient not only plays the primary role in the gravity-induced term of the intruder segregation force, (\ref{eq:fseg1}), but also influences the kinematics-induced term, (\ref{eq:fseg2}). The theoretical lithostatic pressure (solid line) is $p=P_0+\rho \phi g_0(H-z)$, where the solid volume fraction is assumed to be a constant $\phi=0.55$, and the applied overburden pressure is half of the maximum lithostatic pressure, $P_0=0.5\rho \phi g_0H$. The measured DEM pressures (data points), including both the dynamic, which is negligible, and the static components~\citep{luding2008introduction}, match the expected values. For the linear and exponential velocity profiles (columns 1 and 4), gravity is imposed perpendicular to the flow direction.  As a result, the pressure increases linearly with depth from the imposed overburden pressure, $P_0$, applied at the top wall, to $3 P_0$ at the bottom wall due to the added weight of the flowing particles (d1, d4). The resulting pressure gradient, ${\partial p}/{\partial z}$, is constant due to the linear pressure increase with depth. For the two other cases (columns 2 and 3), $g=0$ and the flow volume is constant because the walls are constrained to be $H=0.2$\,m apart.  Consequently, the resulting pressures are constant (see caption) and the pressure gradients in the $z$-direction are zero (d2, d3). Although $g=0$ in these two cases, $P_0$ is expressed relative to Earth's gravity, $g_0=9.81$\,m/s$^2$, to allow comparison with the $g\neq0$ cases and to provide physical context.

The bottom row in figure~\ref{fig3} shows the concentration profile of large particles, $c_l$ (color), and the solid volume fraction profile, $\phi$ (black). In all cases, $c_l=0.5$ (vertical colored line) within the uncertainty except near the walls, where size exclusion effects become significant. The concentration is nearly constant because segregation is suppressed by the restoring force, as described earlier in the context of figure~\ref{fig_fres}. The solid volume fraction shows only minimal variation with depth, and remains near $\phi=0.55$ (vertical black line) in all cases, which is typical for these flow conditions~\citep{jing_rising_2020}.

\begin{figure}
    \centerline{\includegraphics[width=\columnwidth]{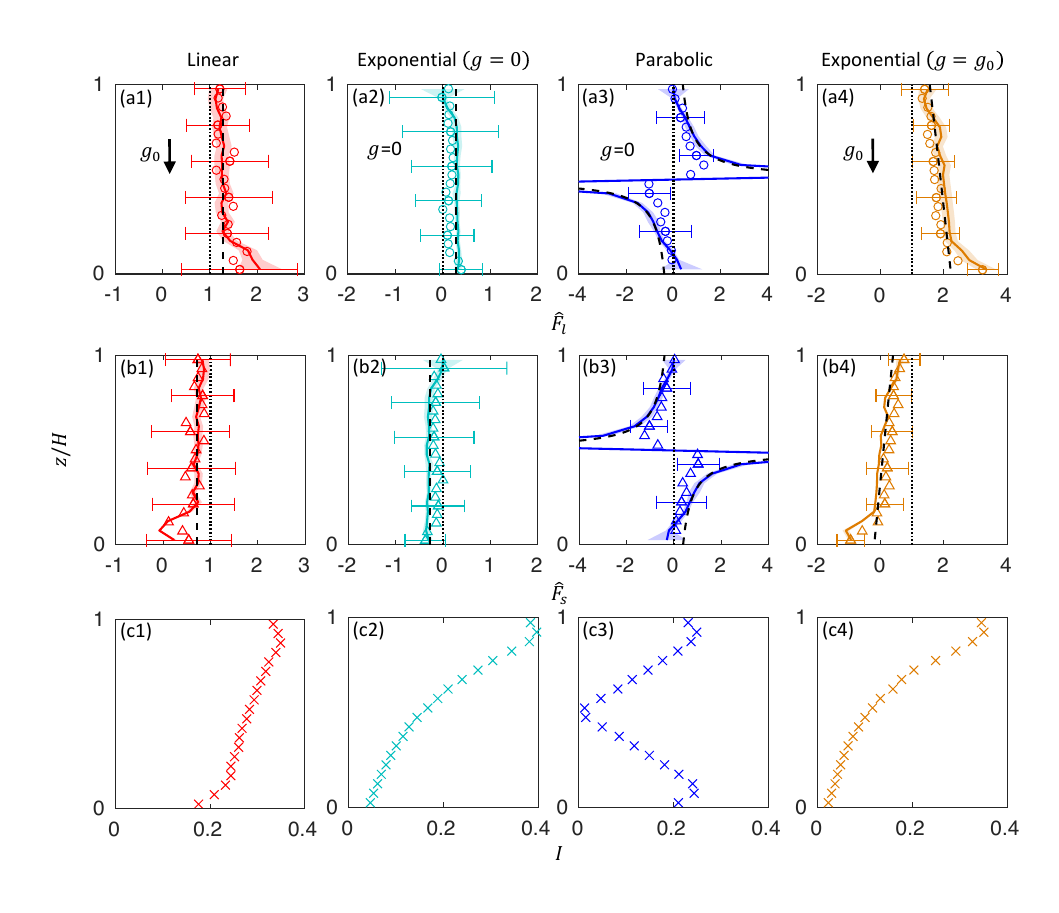}}
\caption{Segregation force profiles for large, $\hat{F}_l={F_l}/m_lg_0$ (row a), and small, $\hat{F}_s={F_s}/m_sg_0$ (row b), particles at $R=2$ from the model (\ref{tanh2}) using the imposed velocity profile (dashed black curves) and the measured profiles (solid color curves) in figure~\ref{fig3} as well as direct DEM measurements (symbols) from 2\,s time averages after the flow reaches steady state. Note the different horizontal axes limits in rows a and b. Vertical dotted lines indicate the value about which $\hat{F}_{l}$ and $\hat{F}_{s}$ should balance according to (\ref{eq:fseg_sum1}). Error bars indicate the standard deviation of the DEM data over the 2\,s sampling window. Shaded bands represent the uncertainty of the segregation force calculated from the measured profiles. (row c) Inertial number profiles, $I=\dot{\gamma} \bar{d} / \sqrt{p/\rho}$ (see text).
}
    \label{fig4}
\end{figure}

With the various flow fields characterized, the intruder segregation force, $\hat{F}_{i,0}$, can be determined and incorporated into the concentration-dependent form for the local segregation force on a particle, $\hat{F}_{i}$. Specifically, $\hat{F}_{i}$, is calculated from the corresponding local values of $\dot\gamma$, $\partial \dot\gamma/\partial z$, $p$, and $\partial p/\partial z$ according to (\ref{eq:fseg}) with coefficients from (\ref{eq:intruder}) and modified per (\ref{tanh2}) to account for the particle concentration. The values of $\dot\gamma$, $\partial \dot\gamma/\partial z$, $p$, and $\partial p/\partial z$ can be based on either their imposed values (solid curves in figure~\ref{fig3}) or their measured DEM values (data points in figure~\ref{fig3}).  Hence, we plot three $\hat{F}_{i}$ results for $R=2$ in figure~\ref{fig4}(a,b): a dashed black curve for $\hat{F}_{i}$ based on the imposed values of $\partial \dot\gamma/\partial z$, $p$, and $\partial p/\partial z$, a colored solid curve for $\hat{F}_{i}$ based on the DEM measurements of $\partial \dot\gamma/\partial z$, $p$, and $\partial p/\partial z$, and data points for the values of $\hat{F}_{i}$ based on direct force measurements from DEM. Error bars indicate the DEM data standard deviation over the 2\,s window, sampled at 0.01\,s intervals (shown only for every fourth data point to avoid obscuring other data). Shaded error bands represent the uncertainty in force prediction, derived from the standard deviations of the time-averaged flow fields that propagate through the force model, $\sigma_F=F \sqrt{ 2(\frac{\sigma_p}{p})^2+ (\frac{\sigma_{c_l}}{c_l})^2 +(\frac{\sigma_{c_s}}{c_s})^2 +(\frac{\sigma_{\dot\gamma}}{\dot\gamma})^2}$. Again, we express $\hat{F}_i={F_i}/m_ig_0$ values relative to $g_0=9.81$\,m/s$^2$, even when the imposed gravitational field is zero to allow comparison with the non-zero gravity cases and to provide physical context. The vertical dotted lines in the first two rows of the figure indicate the value about which $\hat{F}_{l}$ and $\hat{F}_{s}$ should balance according to (\ref{eq:fseg_sum1}), which is $\cos \theta$ for $g\neq 0$ and 0 for $g=0$ (zero gravity component in the segregation direction is equivalent to $\theta=\pi/2$).

Figure~\ref{fig4} shows that, overall, the predicted segregation forces, $\hat{F}_l$ in row (a) for large particles and $\hat{F}_s$ in row (b) for small particles, match the DEM data for all cases, regardless of whether the prediction is based on the imposed velocity profile (dashed black curves) or the measured profiles (solid color curves). The good match is unsurprising for the uniform shear flow as this is the gravity-induced segregation case upon which the concentration dependence in (\ref{eq:fseg_g}) is based. However, the strong agreement in the other three cases demonstrates the validity of our approach. 

In detail, first consider the linear velocity profile in column~1 of figure~\ref{fig4}. Here the kinematics-induced segregation is zero and all segregation is due to gravity, for which $\partial p / \partial z$ is constant. Hence, the segregation forces, $\hat{F}_l$ and $\hat{F}_s$, are constant with depth. More importantly, $\hat{F}_{i}$ based on the imposed values of $\partial p/\partial z$ (dashed black lines), $\hat{F}_{i}$ based on the DEM measurements of $\partial p/\partial z$ (solid color curves), and $\hat{F}_{i}$ based on direct force measurements (data points) match quite well.  Furthermore, $\hat{F}_l>1$, indicating an upward segregation force on large particles due to the pressure gradient force that exceeds the particle weight, resulting in a tendency for unconstrained large particles to rise. Commensurately, $\hat{F}_s<1$, indicating a small-particle segregation force that is less than the particle weight, resulting in a tendency for small particles to sink. Of course, due to the imposed restoring forces (see figure~\ref{fig_fres}), the initial mixed concentration profile, $c_l=c_s=0.5$ remains uniform throughout the domain. Also note that the concentration weighted sum of $\hat{F}_l$ and $\hat{F}_s$ is one. In other words, the total segregation force across both species for the entire system sums to the total particle weight, as indicated by (\ref{eq:fseg_sum1}). This is evident in figure~\ref{fig4}, column 1, as $\hat{F}_l$ and $\hat{F}_s$ being equidistant on either side of the dashed vertical line at $\hat{F}_i=1$ for both the DEM measurements and the model predictions.

The match between the model predictions and the DEM data for the exponential velocity profile in column 2 of figure~\ref{fig4}, while imperfect, indicates that kinematics-induced segregation can be captured by the extension of the intruder particle segregation force in (\ref{eq:fseg}) using the concentration dependence described by (\ref{tanh2}). In this case, $\hat{F}_l$ and $\hat{F}_s$ depend only on the kinematics-induced term in (\ref{eq:fseg2}) to which $\dot\gamma$, $\partial \dot\gamma/\partial z$, and $p$ all contribute. However, even though $\dot\gamma$ and $\partial \dot\gamma/\partial z$ vary with depth for the exponential profile (see figure~\ref{fig3} (b2, c2)), the product $(1/\dot\gamma) \partial \dot\gamma/\partial z$ is constant, as is $p$. Hence, the kinematics-induced segregation force is depth independent. The model predictions based on both the imposed velocity profile (dashed black line) and the measured velocity (color solid curve) slightly overestimate the magnitude of $\hat{F}_l$ and $\hat{F}_s$. The model's underestimate of the segregation forces does not appear to be related to issues with the profiles in figure~\ref{fig3}, for which the measured profiles for $\dot\gamma$, $\partial \dot\gamma/\partial z$, and $p$ seem to follow the imposed profiles quite closely. Nevertheless, $\hat{F}_l>0$, indicating a segregation force due to the shear that is in the positive $z$-direction for the large particles, reflecting the tendency of large particles to segregate toward regions of higher shear rate in dense flows~\citep{fan_theory_2011,jing_unified_2021}. In contrast, $\hat{F}_s<0$, which reflects the tendency of small particles to segregate toward low shear regions. Additionally, because $g=0$ in this case, the total segregation force sums to 0 instead of 1 ((\ref{eq:fseg_sum1}) becomes $ c_l \hat{F}_l + c_s \hat{F}_s=0$), as is evident in figure~\ref{fig4}(a2, b2) and verified mathematically from the data.

The match between the DEM data and the model predictions for the parabolic velocity profile is more convincing of the model's validity. In this case, the pressure gradient is zero, so the segregation force is again entirely kinematics-induced. However, there are two complications in the kinematics-induced term at $z/H=0.5$. First, there is a singularity in the kinematics-induced term in (\ref{eq:fseg2}), because $\dot\gamma = 0$, and, second, the segregation force switches sign. Both effects are evident in the measured segregation force and the model predictions. Given this situation, it is not surprising that there is substantial deviation of the predictions from the measured segregation force around $z/H=0.5$. While the model predicts the strong curvature in the dependence of the segregation forces on $z$, it again overpredicts the segregation forces compared to DEM measurements. The exception is near the walls where the DEM measurements lie between the model prediction determined using the imposed velocity profile (dashed black curve) and the model prediction determined using the measured velocity profile (solid color curve). Clearly, the model captures the qualitative dependence of the segregation force on the local kinematics, although the quantitative agreement could be better. Again, since $g=0$, $ c_l \hat{F}_l + c_s \hat{F}_s=0$, as is evident in figure~\ref{fig4}(a3, b3) and verified mathematically from the data.

The last controlled shear flow that we consider combines gravity-induced and kinematics-induced segregation using an exponential velocity profile with gravity, see column 4 of figure~\ref{fig4}. Here the combined effects of the pressure and shear rate gradients result in a linear dependence of $\hat{F}_i$ on $z$.  The upward segregation force on large particles increases with depth, while the segregation force on small particles decreases with depth to the point of changing from positive to negative near $z/h \approx 0.2$. Nevertheless, $\hat{F}_l$ and $\hat{F}_s$ are equidistant on either side of the dotted vertical line at $\hat{F}_i=1$, indicating that (\ref{eq:fseg_sum1}) is satisfied. The match between the model predictions and the DEM measurements of the segregation force are reasonable. It is also evident that the model predictions based on the measured velocity profile capture a portion of the impact of the lower wall on the segregation force.

To further assess the applicability of the segregation force model proposed here, we also plot profiles of the local inertial number, $I=\dot{\gamma} \bar{d} / \sqrt{p/\rho}$ where $\bar{d}=\sum c_id_i$, in the bottom row of figure~\ref{fig4}. It is evident that with $I$ ranging from almost zero to nearly 0.4, a wide range of inertial numbers consistent with dense granular flows are represented by the cases considered here.

\begin{figure}
\centerline{\includegraphics[width=\columnwidth]{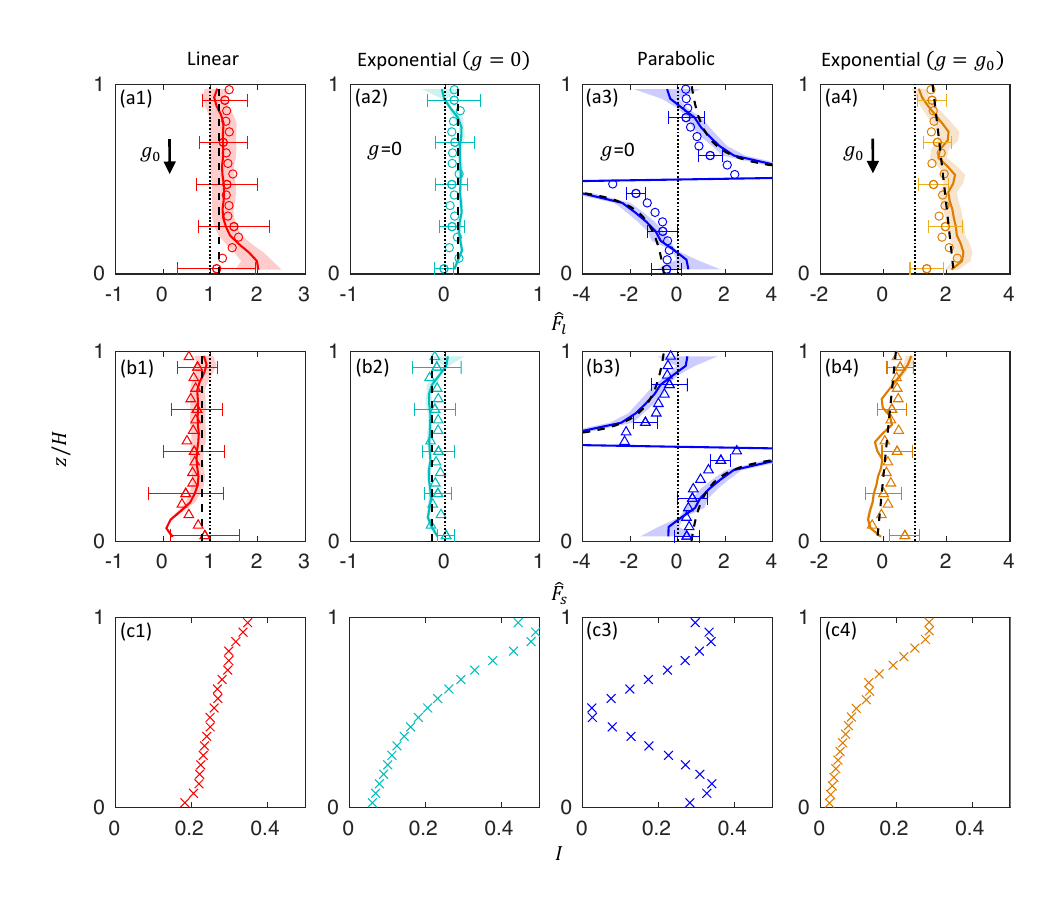}}
\caption{Segregation force profiles for large, $\hat{F}_l={F_l}/m_lg_0$ (row a), and small, $\hat{F}_s={F_s}/m_sg_0$ (row b), particles at $R=3$ ($d_l=4$\,mm, $d_s=4/3$\,mm, $\rho_l=\rho_s=1$\,g/cm$^3$, $H\approx 0.2$\,m$=50d_l$, and $U=20$\,m/s) from the model (\ref{tanh2}) using the imposed velocity profile (dashed black curves) and the measured profiles (solid color curves) in figure~\ref{fig3} as well as direct DEM measurements (symbols) from 2\,s time averages after the flow reaches steady state. 
Note the different horizontal axes limits in rows a and b.
Vertical dotted lines indicate the value about which $\hat{F}_{l}$ and $\hat{F}_{s}$ should balance according to (\ref{eq:fseg_sum1}). Error bars indicate the DEM data standard deviation over the 2\,s sampling window. Shaded bands represent the uncertainty of the segregation force calculated from the measured profiles. (row c) Inertial number profiles, $I$.
} 
\label{fig4_R3}
\end{figure}

At this point we return to (\ref{eq:intruder}), which is an empirical fit to data for which there is substantial scatter, particularly for $1\lesssim R\lesssim 2.5$, as well as some inertial number dependence~\citep{jing_unified_2021}, either of which could explain the systematic offset of our DEM data from the model predictions for $\hat{F}_i$ in columns 2-4 of figure~\ref{fig4}.  Hence, we additionally evaluate the concentration-dependent segregation force model for the case of $R=3$ ($d_l=4$\,mm, $d_s=4/3$\,mm), where these effects are smaller (though still not insignificant). Again, values for $f^g$ and $f^k$ come from (\ref{eq:intruder}), as described earlier. Results for the segregation force are shown in figure~\ref{fig4_R3}, where the flow conditions are identical to those described in the caption of figure~\ref{fig3}, except that $R=3$. Under these conditions, the pressures that arise from the constant domain volume constraint when $g=0$ are $P_0\approx0.22\rho \phi g_0H$ for the exponential profile and $P_0\approx0.99\rho \phi g_0H$ for the parabolic profile.

It is immediately evident that the quality of the match between the segregation force model predictions and the DEM measurements of the segregation force profiles is similar for $R=2$ and $R=3$. The model, whether using imposed profiles for $\dot\gamma$, $\partial \dot\gamma/\partial z$, $p$, and $\partial p/\partial z$ or measured profiles for these quantities, generally coincide. Perhaps more importantly, the measured segregation force obtained directly from the DEM results matches both models reasonably well for $R=3$. Of course, the standard deviation of the DEM data is large compared to the values of $\hat{F}_i$, and the agreement is not as good near the walls and around the singularity at $z/H=0.5$ for the parabolic case. However, it is clear that the model for concentration dependence of the segregation force, (\ref{tanh2}), combined with the gravity-induced and kinematics-induced terms in the single intruder segregation force model, (\ref{eq:fseg}), agrees reasonably well with the DEM measurements over a range of inertial numbers that are consistent with dense granular flow.

\subsection{Varying concentration}
\label{sec:concentration}
\begin{figure}
    \centerline{\includegraphics[width=\columnwidth]{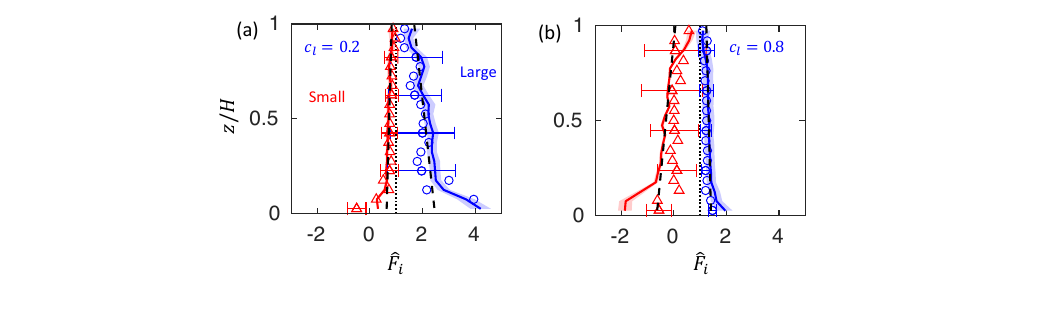}}
\caption{Segregation force profiles for $\hat{F}_l={F_l}/m_lg_0$ (blue circles) and $\hat{F}_s={F_s}/m_sg_0$ (red triangles) with $R=2$ for the exponential velocity profile with bulk large particle concentrations (a) $c_l=0.2$ and (b) $c_l=0.8$, based on the imposed velocity profiles (dashed lines) and the measured profiles (solid curves) compared to DEM measurements (symbols) averaged over 2\,s after the flow reaches steady state.  Vertical dotted lines indicate the value about which $\hat{F}_{l}$ and $\hat{F}_{s}$ should balance according to (\ref{eq:fseg_sum1}). Error bars indicate the DEM data standard deviation over the 2\,s sampling window. Shaded bands represent the uncertainty of the segregation force calculated from the measured profiles. 
}
\label{fig4_concentration}
\end{figure}

The previous section considers only uniform mixtures of equal small and large particle volumes ($c_s=c_l=0.5$). However, the concentration dependence of the segregation force based on (\ref{tanh2}) should be valid for any concentration, $0 \leq c_l \leq 1$ with $c_s=1-c_l,$ and for non-uniform spatial concentration as well. To test this, we consider the exponential velocity profile case of figure~\ref{fig2}(d) because it includes both gravity-induced and kinematics-induced segregation. Model predictions for $\hat{F}_i$ for uniform concentrations of $c_l=0.2$ and $c_l=0.8$ at $R=2$ are shown in figure~\ref{fig4_concentration}. Whether based on imposed profiles for $\dot\gamma$, $\partial \dot\gamma/\partial z$, $p$, and $\partial p/\partial z$ or measured profiles of these same quantities, the model predictions generally coincide with each other as well as with the DEM results, although the measured  $\hat{F}_i$ values for the lower concentration species tend to be closer to zero than the predicted values. This is likely because the the segregation forces are quite noisy (large error bars) and the fit parameters used in (\ref{eq:intruder}) are not entirely accurate. Although it is not evident from figure~\ref{fig4_concentration}, the total segregation force for both $c_l=0.2$ and $c_l=0.8$ sums to one as expected from (\ref{eq:fseg_sum1}). 

Up to this point and in all cases, we start with a uniform concentration of small and large particles in the flow domain, apply a spring-like restoring force to the particles within each layer to maintain the fully mixed condition, and measure the local segregation force for each particle type, as outlined in the context of figure~\ref{fig_fres}. However, the same approach can also be applied when the particle species concentration varies with depth. That is, each layer shown in figure~\ref{fig_fres} would have a different concentration of small and large particles. 
Here, we consider that situation with controlled linear and exponential velocity profiles with gravity, corresponding to the flows shown schematically in figure~\ref{fig2}(a, d).

\begin{figure}
    \centerline{\includegraphics[width=\columnwidth]{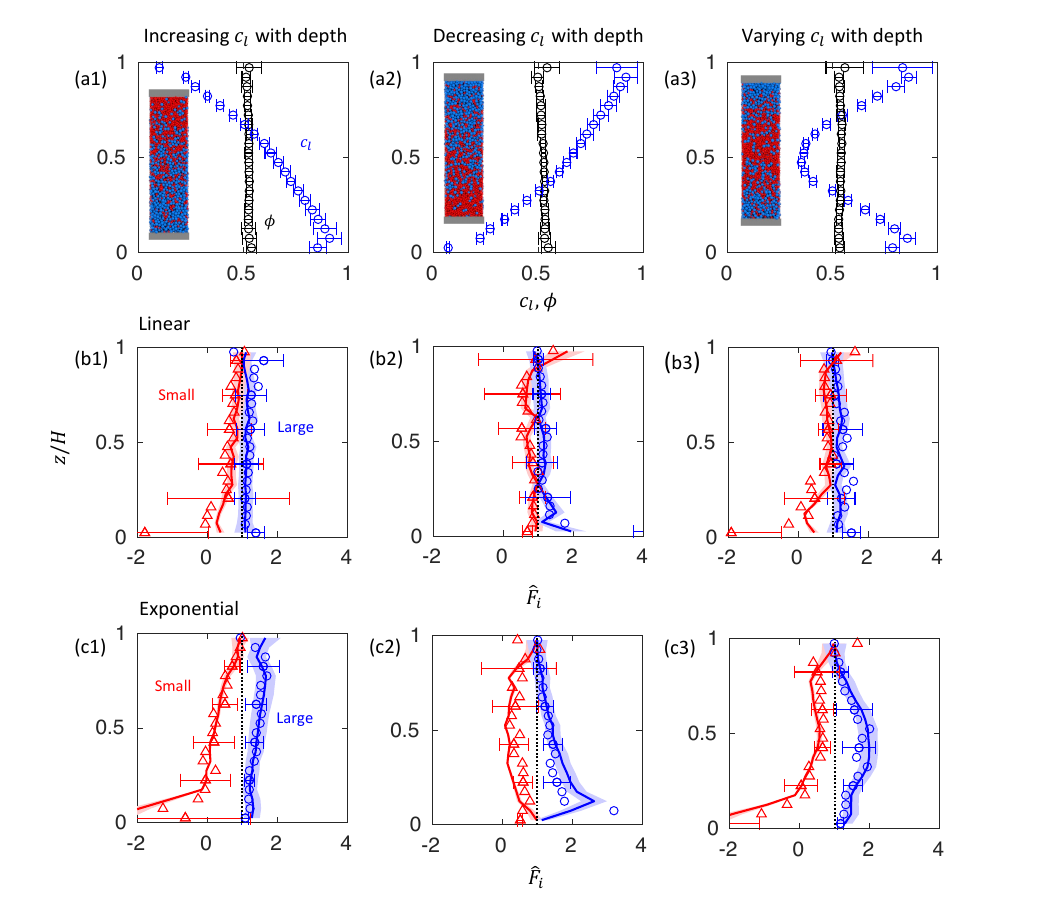}}
\caption{Concentration, solid volume fraction, and segregation force profiles for $\hat{F}_l={F_l}/m_lg_0$ (blue) and $\hat{F}_s={F_s}/m_sg_0$ (red) at $R=2$ for large particle concentrations varying with depth (row a). Segregation force profiles for large (circles) and small (triangles) particles in a controlled uniform shear flow with gravity (row b) and an exponential velocity profile with gravity (row c) with the same conditions as in figure~\ref{fig3}. $\hat{F}_i$ is based on the measured profiles (solid curves) as well as DEM measurements (symbols) averaged over 2\,s after the flow reaches steady state. Vertical dotted lines indicate the value about which $\hat{F}_{l}$ and $\hat{F}_{s}$ should balance when weighted by the concentration according to (\ref{eq:fseg_sum1}). 
Note that the truncated data point in (c3)  near $z/H=0$ with $\hat F_s=-3.12$ matches the model prediction of -2.72 (red curve) within the range of uncertainty.  
}
    \label{fig_varyingc}
\end{figure}

Three large-particle concentration profiles are considered (with $c_s=1-c_l$), as shown in the top row of figure~\ref{fig_varyingc}: increasing $c_l$ with depth, decreasing $c_l$ with depth, and decreasing $c_l$ in the top half of the flow and increasing $c_l$ in the bottom half of the flow. In all three cases, the solid volume fraction $\phi$ is nearly constant from top to bottom. Note that the concentration profiles in the first two cases (a1, a2) are slightly nonlinear due to packing that occurs after initially filling the system with a linearly varying concentration of  particles, and, in all cases, a slight wall exclusion effect is evident in $c_l$, as expected. Also, as a result of the non-trivial dependence of $c_l$ and $c_s$ on $z$, it is not possible to determine the model-predicted values for $\hat{F}_i$ based on the imposed velocity profiles, which are shown as dashed black curves in preceding figures.

When the concentration of large particles increases with depth (column 1), the predicted segregation forces match the measured forces for both the linear (b1) and exponential (c1) velocity profiles. Since $\hat{F}_l>1$, the segregation force is upward and exceeds the particle weight for the large particles, resulting in a tendency for free large particles to rise. Commensurately, $\hat{F}_s<1$, indicating a segregation force on small particles that is less than the particle weight, resulting in a tendency for free small particles to sink. Of course, due to the imposed restoring forces (see figure~\ref{fig_fres}), the initial concentration profile is maintained. Specifically, for the linear velocity profile (b1), $\hat{F}_l$ remains slightly above one through the entire depth, but $\hat{F}_s$ decreases further below one with increasing depth, particularly for small $z/H$. This is consistent with the fact that the segregation velocity of a species increases as its local concentration decreases for most segregation velocity models~\citep{jones_asymmetric_2018} --- as $c_l$ increases with depth, the segregation velocity of small particles increases. However, the segregation force is not restricted to have the same trend as the segregation velocity because the segregation velocity results from the imbalance of all forces acting on a particle including the drag force, which we do not consider here. The segregation forces are larger for the exponential velocity profile (c1). For both velocity profiles, the model predictions based on the measured concentration and flow fields match the DEM measurements of the segregation force. Note that the measured segregation force, $\hat{F}_i$, has large uncertainty when $c_i$ is small due to the small number of associated particles available for averaging at low concentrations.

When the large particle concentration decreases with depth (column 2), again $\hat{F}_l>1$, indicating an upward segregation force that exceeds the particle weight, and $\hat{F}_l$ increases deeper in the bed where $c_l$ is smaller, particularly for the exponential velocity profile (c2). As in column 1, $\hat{F}_s<1$, indicating that the segregation force for small particles is less than the particle weight, indicating a tendency for small particles to sink. The model prediction matches the DEM measurement reasonably well for both the linear and exponential velocity profiles. Similarly for the $c_l$ profile with a minimum at $z/H=0.5$ (column 3), the model prediction matches the measured segregation force reasonably well, even with large changes in the concentration gradient. For all cases in figure~\ref{fig_varyingc}, (\ref{eq:fseg_sum1}) is satisfied locally. 

Unlike flows with uniform particle concentrations, particles in flows with depth-varying concentration profiles tend to diffuse toward a uniform concentration state in the absence of segregation forces. This diffusive flux resulting from the concentration gradients is balanced by the force-measurement-imposed restoring force and the segregation force, ultimately affecting the measured segregation force.
However, the fact that the segregation force model (\ref{tanh2}) does not specifically account for this diffusive effect and yet still predicts the measured segregation force reasonably well indicates that diffusion-driven offsets in species centers of mass in the measurement layers of the flow (see figure~\ref{fig_fres}) due to  concentration gradients have negligible impact for the cases we consider here.

Overall, it is evident that the concentration dependent segregation force model (\ref{tanh2}), which relies on the intruder segregation force based model (\ref{eq:fseg}), can estimate segregation even when the concentration fields are spatially varying.  Additionally, this prediction capability implies that the segregation force is relatively insensitive to concentration gradients.

\begin{figure}  \centerline{\includegraphics[width=\columnwidth]{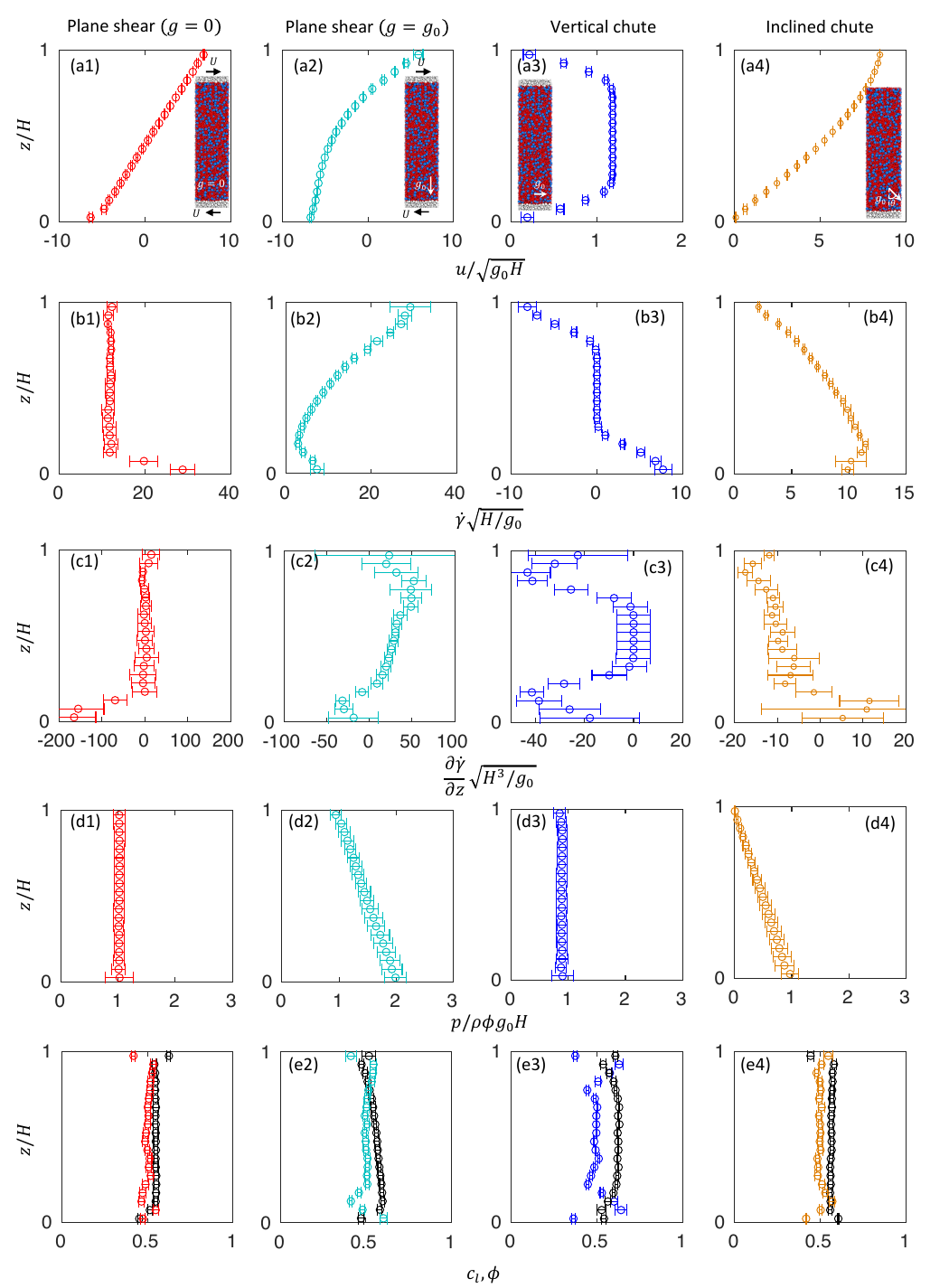}}
\caption{Scaled flow fields of natural shear flows with $R=1.5$, $d_l=4$\,mm, $d_s=8/3$\,mm, $\rho_l=\rho_s=1$\,g/cm$^3$.  Left to right: plane shear without and with gravity, vertical chute, and inclined chute. $H\approx0.2\,\mathrm{m}=50d_l$ for wall-bounded cases  and fixed $H=0.2$~m for vertical chute case, $g=g_0=9.81$\,m/s$^2$, and $\theta=28^{\circ}$ for the inclined chute. Applied overburden pressure $P_0=\rho \phi g_0H$ (columns 1 and 2).  Depth-averaged pressure for the vertical chute $P_0=0.89\rho \phi g_0H$  (column 3). (row e) $c_l$ data symbols are colored, and $\phi$ data symbols are black.
}
    \label{fig5}
\end{figure}

\subsection{Natural shear flows}
\label{sec:natural_shear}
As demonstrated above, the intruder segregation force model (\ref{eq:fseg}) with gravity- and kinematics-driven terms can be extended to apparently arbitrary concentrations and concentration fields via (\ref{tanh2}) in a variety of flows where the velocity field is artificially controlled.  We now examine four uncontrolled wall- or gravity-driven flows illustrated in figure~\ref{fig2}(e-h) in which the velocity field develops naturally via the boundary conditions and gravity-induced body forces. Of note in the three examples with gravity (f-h) is the direction of gravity with respect to the flow, which is characterized by the angle of the bottom wall with respect to horizontal, $\theta$. For wall-driven flow, $g$ is perpendicular to the flow direction (in $z$), such that $\theta=0$; for inclined chute flow, $\theta$ is greater or equal to the critical angle for flow to occur; % \geq \theta_{crit}$, where $\theta_{crit}$ is a critical chute angle for flow to occur;
and for vertical chute flow $g$ is parallel to $z$, such that $\theta=\pi /2$ and $\partial P/\partial z=0$. 

As in the analysis of the controlled-velocity flow fields in the previous sections, we first plot dimensionless depth profiles of $u$, $\dot\gamma$, $\partial \dot\gamma/\partial z$, $p$, $c_l$, and $\phi$ with $c_l=c_s=0.5$ for the natural flows, as shown in figure~\ref{fig5}. The kinematic-terms are scaled with $g_0$ and $H$, since there is no intrinsic velocity scale for the vertical- or inclined-chute cases. We use $R=1.5$ here to test a third size ratio (similar results for these natural flows are achieved for other size ratios). 

Consider first the wall-driven flow without gravity (column 1). The velocity profile is nearly linear with depth, except for a slight deviation near the lower wall which is amplified for $\dot\gamma$ and $\partial \dot\gamma/\partial z$. The slightly asymmetric velocity profile near the top and bottom walls is due to the top wall being able to move vertically to accommodate dilation during flow, while the bottom wall is fixed vertically. The profiles of pressure and solid volume fraction, $\phi$, are nearly constant, while the concentration profile shows small deviations from its mean value near the walls. With gravity (column 2), the wall-driven flow velocity profile is steep near the upper moving wall at $z/H=1$ but flattens in the bottom half of the flow where the pressure is higher. This results in $\dot\gamma$ and $\partial \dot\gamma/\partial z$ decreasing near the bottom of the flow. At the same time the pressure increases linearly with depth, and the pressure gradient is nearly constant. There is a small increase in $\phi$ with depth as particles near the bottom wall dilate less due to the larger local overburden pressure. 

The vertical chute flow (column 3) has a plug-like velocity profile, resulting in $\dot\gamma$ varying from negative to positive with depth, while $\partial \dot\gamma/\partial z \leq 0$ with widely varying values. The pressure remains nearly constant at $P_0 \approx 0.89\rho \phi g_0 H$, while the solid volume fraction decreases near the walls compared to the centre of the chute, as observed previously~\citep{fan_theory_2011}. Despite the restoring force to prevent segregation, $c_l$ varies somewhat in the region where $\dot{\gamma}$ is nonzero. Finally, the curvature of the velocity profile for the inclined chute with $\theta=28^{\circ}$ (column 4) is opposite that of the wall-driven flow with gravity (column 2). Consequently, $\dot\gamma$ increases with depth, while $\partial \dot\gamma/\partial z$ is negative, except near the bottom wall, and relatively small through most of the depth compared to the other two flows with gravity. Like the wall-driven case with gravity, $p$ increases with depth, but is zero at the free surface, and $\rho g H \cos{\theta} < \rho g H$ at the base of the flow. The solid volume fraction is independent of depth. 

\begin{figure}    \centerline{\includegraphics[width=\columnwidth]{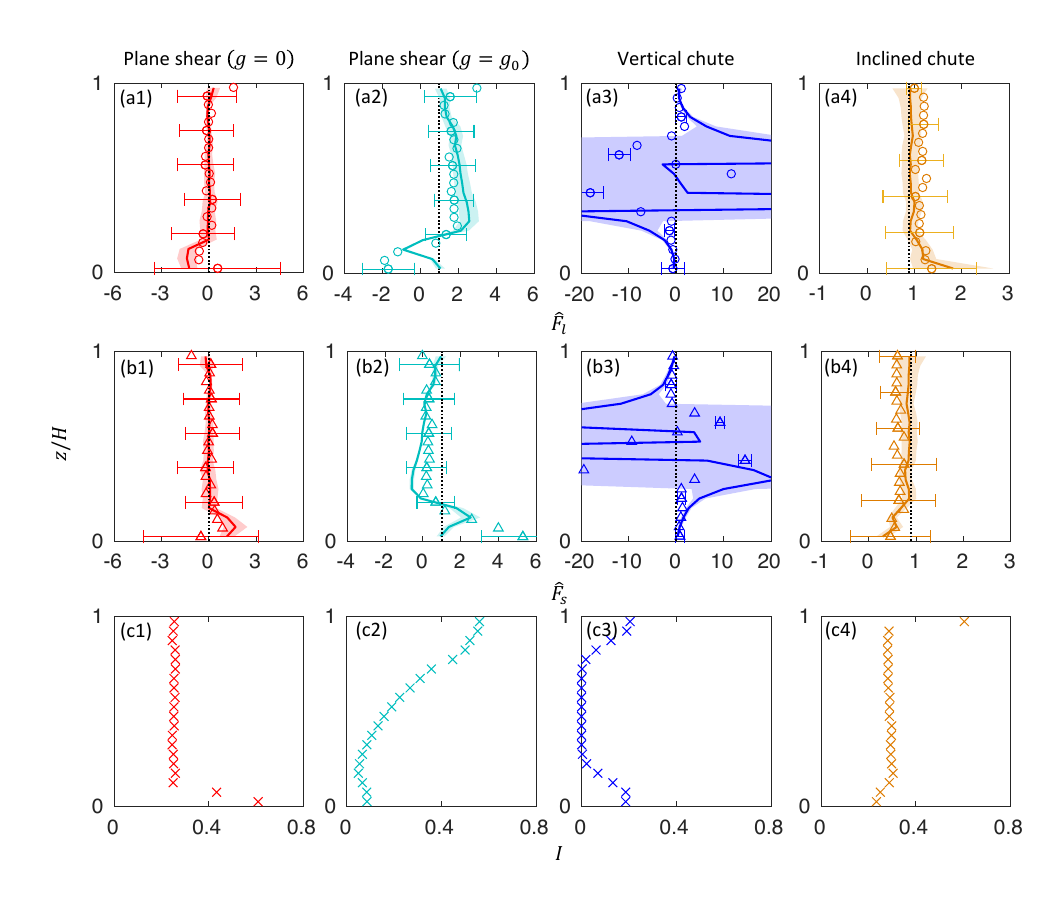}}
\caption{Segregation force profiles for $\hat{F_l}=F_{l}/m_lg_0$ (row a) and $\hat{F_s}=F_{l}/m_sg_0$ (row b) in natural shear flows for $c_l=c_s=0.5$ and $R=1.5$ based on model predictions using measured flow fields in figure~\ref{fig5} (solid color curves) and DEM measurements (symbols) time averaged over 2\,s after the flows reach steady state.
Note the different horizontal axes limits in rows 1 and 2.
Error bars indicate the standard deviation over the 2\,s averaging interval. Shaded bands represent the uncertainty of the segregation force calculated from the measured profiles. (row c) Inertial number profiles, $I$.
}
    \label{fig6}
\end{figure}

Model predictions of $\hat{F}_i$ for the four natural shear flows are shown in figure~\ref{fig6}. Unlike the flows with controlled velocity profiles, these predictions are based only on flow profiles calculated from the DEM simulations (solid color curves), since there is no imposed velocity profile to consider. For the wall-driven shear flow without gravity (column 1), the DEM measured forces match the nominal value of zero and the model predictions, except near the bottom wall where $\hat{F}_l$ is negative and $\hat{F}_s$ is positive for the DEM measurements due to wall effects evident in column 1 of figure~\ref{fig5}. However, it is notable that the model reflects the measured nonzero forces at the bottom wall reasonably well. For the wall-driven shear flow with gravity (column 2), the DEM measured forces and the model prediction show the same trends---increasing with depth in the upper portion of the flow and then decreasing, changing sign, and reaching a relatively larger amplitude near the lower wall. The match is not as good in the lower portion of the flow as in the upper portion. 

For the  vertical chute flow (column 3), the model-predicted forces match the DEM measurements near the walls (within $0.2H$), which is where the shear rate is non-zero and the inertial number is not too small (c3). In the central portion of the chute where the flow is plug-like, $\dot\gamma \approx 0$, $\partial \dot\gamma/\partial z \approx 0$, and $I \approx 0$, the model correctly predicts the change in the direction of the segregation force from $z/H \approx 0.3$ to $z/H \approx 0.7$, but with extremely large uncertainty between these two heights and with much larger predicted forces than the DEM measurements.  This large deviation is likely due to a breakdown of the model validity due to the corresponding very low inertial number (c3) and the singularity in the kinematics term in (\ref{eq:fseg2}) associated with $\dot\gamma \to 0$. Lastly, the model prediction show the same trends as the DEM measurements for the inclined chute flow (column 4). The model underestimates the measured segregation forces for $z/H>0.2$ where the measured segregation forces are small and noisy, but matches the data well near the bottom wall ($z/H<0.2$) where $\hat{F_l}$ and $\hat{F_s}$ are largest. In all cases, the sum of the segregation force across the two species is satisfied per (\ref{eq:fseg_sum1}).

While the correspondence between the model predictions and DEM measurements of the segregation forces in these four natural flows is somewhat less satisfying than that for the controlled flows, these results nevertheless demonstrate that the intruder force model of (\ref{eq:fseg}) can be applied to bidisperse mixtures with reasonable accuracy using (\ref{eq:intruder}) and (\ref{tanh2}), except in regions where $\dot\gamma \approx 0$.

\section{Conclusions}
Predicting the segregation force on single intruder particles, not to mention the more difficult problem of particles in mixtures, in granular flows confounded researchers for decades until the virtual spring approach pioneered by~\cite{guillard_scaling_2016} allowed it to be directly measured. Using that method, we established the dependence of the segregation force on gravity and local kinematics for an intruder particle in three-dimensional granular flow of spherical particles via (\ref{eq:fseg}) and (\ref{eq:intruder})~\citep{jing_unified_2021}. We then extended the virtual spring measurement method to allow measurement of segregation forces in finite-concentration size-bidisperse mixtures with pressure gradients via (\ref{eq:fseg_g})~\citep{duan_segregation_2022}. Here, we further extend the model for the combined gravity- and kinematics-induced segregation force on an intruder particle (\ref{eq:fseg}) to arbitrary concentrations of size-bidisperse particle mixtures, (\ref{tanh2}), by applying the concentration dependence described by (\ref{eq:fseg_g}) to both gravity- and kinematics-induced components of the segregation force.  We use an extensive set of DEM simulations to show that the approach can estimate the segregation force in four idealized flows with an artificially controlled velocity profile as well as four natural shear- and gravity-driven flows (subject to minor deviations near walls for wall-driven flows and for very small inertial numbers). 

Considering the dependence of the segregation force on the particle size ratio, the velocity field and its gradients, the pressure field and its gradients, and the relative concentrations of the two particle species, the performance of the model, i.e., (\ref{eq:fseg}), (\ref{eq:intruder}), and (\ref{tanh2}), is remarkable. In all the situations that we consider, it is possible to estimate the local spatial- and concentration-dependent segregation force on small and large particles starting only with a knowledge of particle size ratio and the flow conditions, of which the latter can be based on theory or measurement. 

Given the broad range of conditions considered here, this approach is likely generally applicable to a wide range of size-bidisperse granular flows at inertial numbers typical of dense flows. In fact, the success of the combined models of (\ref{eq:fseg}), (\ref{eq:intruder}), and (\ref{tanh2}) demonstrated here for size-bidisperse flows suggests that it is possible to extend the approach to polydisperse and density disperse flows, as well as combined size and density segregation, particularly since the effects of both particle size and density are accounted for by the intruder segregation force \citep{jing_rising_2020,jing_unified_2021}.

The accurate predictions of the segregation force model detailed in this paper are an important piece in the puzzle of predicting segregation in size-disperse granular flows. Although the overall segregation under many conditions can be predicted via continuum models~\citep{schlick_continuum_2016, xiao_modelling_2016, duan_modelling_2021}, accuracy is predicated on knowledge of the dependence of the segregation velocity for each species as a function of relative size or density~\citep{umbanhowar_modeling_2019}. In addition to the segregation force considered here, another important piece needed for predicting the segregation velocity is the drag force on an intruder particle moving through sheared granular beds, which was recently shown to be Stokes-like across a wide range of conditions~\citep{jing_drag_2022}. A simple force balance on an intruder particle incorporating both the segregation force and the drag force allows prediction of the segregation velocity, which is a crucial element in continuum models for predicting overall segregation in granular flows~\citep{umbanhowar_modeling_2019}. This is clearly an appropriate direction for further research.
\\

\section*{Acknowledgments} We are grateful for helpful discussions with Yi Fan and John Hecht. This material is based upon work supported by the National Science Foundation under Grant No. CBET-1929265.

\section*{Declaration of interests} The authors report no conflict of interest.

\bibliographystyle{jfm}
\bibliography{jfm-instructions}

\end{document}